%% file: syritsyn-lat14-proc.tex
\documentclass{PoS}

\usepackage{amsmath}
\usepackage{slashed}
\usepackage{graphicx}

\include{defs}

\title{Initial nucleon structure results with chiral quarks at the physical point}

\ShortTitle{Initial nucleon structure results with physical chiral quarks}

\author{
        \speaker{S.~Syritsyn}~$^a$,
        T.~Blum~$^{a,b}$,
        M.~Engelhardt~$^c$,
        J.~Green~$^d$,
        T.~Izubuchi~$^a$,
        C.~Jung~$^a$,
        S.~Krieg~$^e$,
        M.~Lin~$^f$,
        S.~Meinel~$^{a,g}$,
        J.~Negele~$^h$,
        S.~Ohta~$^{i,j,a}$,
        A.~Pochinsky~$^h$,
        E.~Shintani~$^{a,k}$
        (RBC and LHP collaborations) \\
        \llap{$^a$} RIKEN/BNL Research Center, Brookhaven National Laboratory, Upton, NY, 11973, USA\\
        \llap{$^b$} Physics Department, University of Connecticut, Storrs, CT 06269, USA\\
        \llap{$^c$} Department of Physics, New Mexico State University, Las Cruces, NM 88003,
                    USA\\
        \llap{$^d$} Institut f\"ur Kernphysik, Johannes Gutenberg-Universit\"at Mainz, 
                    D-55099 Mainz, Germany\\
        \llap{$^e$} Bergische Universit\"at Wuppertal, D-42119 Wuppertal, Germany and \\
                    IAS, J\"ulich Supercomputing Centre, Forschungszentrum J\"ulich, D-52425
                    J\"ulich, Germany \\
        \llap{$^f$} Computational Science Center, Brookhaven National Laboratory, Upton, 
                    NY 11973, USA\\
        \llap{$^g$} Department of Physics, University of Arizona, Tucson, AZ 85721, USA\\
        \llap{$^h$} Massachusetts Institute of Technology, Cambridge, MA 02139, USA \\
        \llap{$^i$} Institute of Particle and Nuclear Studies, KEK, Tsukuba, Ibaraki 3050801,
                    Japan \\
        \llap{$^j$} Department of Particle and Nuclear Physics, SOKENDAI, Hayama, Kanagawa, 
                    2400193, Japan \\
        \llap{$^k$} PRISMA Cluster of Excellence, Institut f\"ur Kernphysik and Helmholtz
                    Institute Mainz, Johannes Gutenberg-Universit\"at Mainz, D-55099 Mainz,
                    Germany\\
        E-mail: \email{ssyritsyn@quark.phy.bnl.gov}}

\abstract{
We report initial nucleon structure results computed on lattices with 2+1
dynamical M\"obius domain wall fermions at the physical point generated by the RBC and UKQCD
collaborations. 
At this stage, we evaluate only connected quark contributions. 
In particular, we discuss the nucleon vector and axial-vector form factors, nucleon axial charge
and the isovector quark momentum fraction. 
From currently available statistics, we estimate the stochastic accuracy of the determination
of $g_A$ and $\la x\ra_{u-d}$ to be around 10\%, and we expect to reduce that to 5\% within the next year.
To reduce the computational cost of our calculations, we extensively use acceleration techniques 
such as low-eigenmode deflation and all-mode-averaging (AMA). We present a method for choosing optimal
AMA parameters.
}

\FullConference{The 32nd International Symposium on Lattice Field Theory,\\
		23-28 June, 2014\\
		Columbia University New York, NY}

\begin{document}

%%%%%%%%%%%%%%%%%%%%%%%%%%%%%%%%%%%%%%%%%%%%%%%%%%%%%%%%%%%%%%%%%%%%%%%%%%%%%%%
\section{Introduction}

Exploring nucleon structure in lattice QCD at the physical point became possible in the recent 
few years, and many collaborations have started calculations that no longer require chiral 
extrapolations.
This is a major step forward because it eliminates one of the major sources 
of systematic uncertainty that made difficult validating methods and achieving 
high precision in lattice QCD.

Nucleon structure calculations at the physical point are extremely demanding.
This usually justifies choosing an affordable fermion action such as Wilson~\cite{Green:2012ud}
or twisted mass~\cite{Alexandrou:2013jsa}, even though they do not respect chiral symmetry of
QCD.
However, a number of nuclear and high energy physics problems, e.g., proton decay and neutron
oscillations, require chiral symmetry for computing corresponding nucleon matrix elements, 
and one has to use substantially more expensive domain wall or overlap fermion actions.
Studying nucleon structure, the nucleon vector and axial vector form factors in particular, 
with these actions is therefore a necessary stepping stone for
developing methods to facilitate calculations that will preserve chiral symmetry.

In this report, we show our initial results from computing nucleon structure directly at the
physical point using gauge configurations generated with the $N_f=2+1$ dynamical domain wall fermion
action~\cite{Blum:2014tka}.
We use various improvement techniques, which we discuss in Sec.~\ref{sec:methodology}, and
present initial, low-statistics results for the nucleon form factors, axial charge and 
quark momentum fraction in Sec.~\ref{sec:results}.

%%%%%%%%%%%%%%%%%%%%%%%%%%%%%%%%%%%%%%%%%%%%%%%%%%%%%%%%%%%%%%%%%%%%%%%%%%%%%%%
\section{Methodology\label{sec:methodology}}

For this initial study with chiral quarks at the physical point, we use one ensemble of gauge
configurations generated by the RBC/UKQCD collaborations
with the inverse lattice spacing $a^{-1}=1.730(4)\text{ GeV}$ and the pion mass 
$m_\pi=139.2(4)\text{ MeV}$~\cite{Blum:2014tka}.
This ensemble is generated with the Iwasaki gauge action and the M\"obius Domain Wall
fermion (MDWF) action for $N_f=2+1$ quarks.
The M\"obius formulation~\cite{Brower:2004xi} of chiral fermions allows one to shorten the fifth
(``flavor'') dimension of the traditional domain wall fermion action without increasing 
the quark residual mass.
The lattice size is $48^3\times96$, corresponding to $m_\pi L=3.86$, which should be
sufficient to substantially suppress finite volume effects, as known from the meson sector.

The main challenge in computing hadron observables with domain wall-like actions is the cost of
calculating light quark propagators, especially at the physical point.
We use two methods to accelerate our calculations: acceleration of the conjugate gradient algorithm 
(CG) with low-eigenmode deflation 
%(applied internally in the solver code to the square of the 4D even-odd preconditioned operator) 
and improved stochastic sampling, or \textit{All-mode Averaging (AMA)}~\cite{Blum:2012uh}.
The idea behind AMA is to compute cheap approximate samples with large statistics
and then correct for any potential bias by comparing a subset of approximate samples to exact
solutions,
\begin{gather}
\la\mcO\ra_\text{imp} 
  = \la\mcO_\text{approx}\ra_{N_\text{approx}}
    + \la\Delta\mcO\ra_{N_\text{exact}}\,,
\quad \Delta\mcO = \mcO_\text{exact} - \mcO_\text{approx}\,,
\\
\big(\delta\mcO_\text{imp}\big)^2 =
    \frac1{N_\text{approx}}\text{Var}\big\{\mcO_\text{approx}\big\}
  + \frac1{N_\text{exact}}\text{Var}\big\{\Delta\mcO\big\}
\end{gather}
One usually takes a large number of approximate samples at different locations 
on the same lattice configuration, exploiting the translational invariance of the QCD ensemble 
average, while for $\Delta\mcO$ smaller statistics is sufficient to estimate the difference 
arising from the approximation.
We obtain approximate samples using quark propagators computed with truncated CG.
Low-eigenmode deflation is applied in computing both exact and approximate samples.
In addition to accelerating the CG convergence for the exact samples, deflation helps to reduce 
the bias of approximate samples and make them cheaper by substantially reducing the number 
$n_\text{CG}$ of CG iterations required for good approximation.

\begin{figure}
  \centering
  \begin{minipage}[T]{.40\textwidth}
    \centering
    \includegraphics[width=\textwidth]{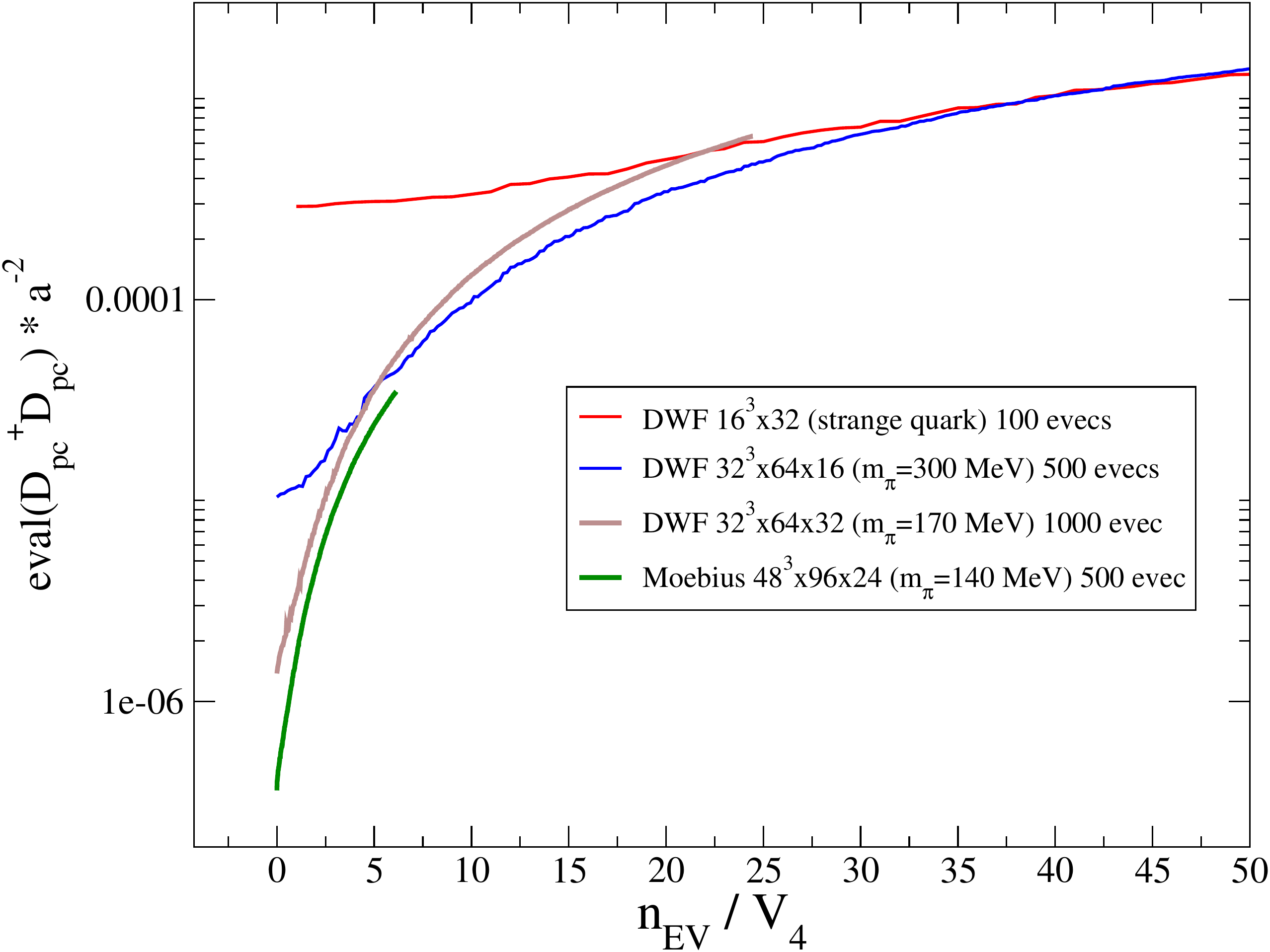}
    \caption{\label{fig:ev-comparison}Deflation eigenvalues for different domain wall operators, 
      scaled with volume and lattice cutoff (courtesy of T.~Blum, T.~Izubuchi, and E.~Shintani).
      The green line corresponds to the current work.}
  \end{minipage}~
  \hspace{12pt}~
  \begin{minipage}[T]{.55\textwidth}
    \centering
    \begin{minipage}[T]{.05\textwidth}
      \rotatebox{90}{$\text{Cost}_\text{imp}\cdot\text{Var}\big\{\mcO_\text{imp}\big\}$}~
    \end{minipage}~
    \begin{minipage}[T]{.47\textwidth}
      \centering
      \includegraphics[width=\textwidth]{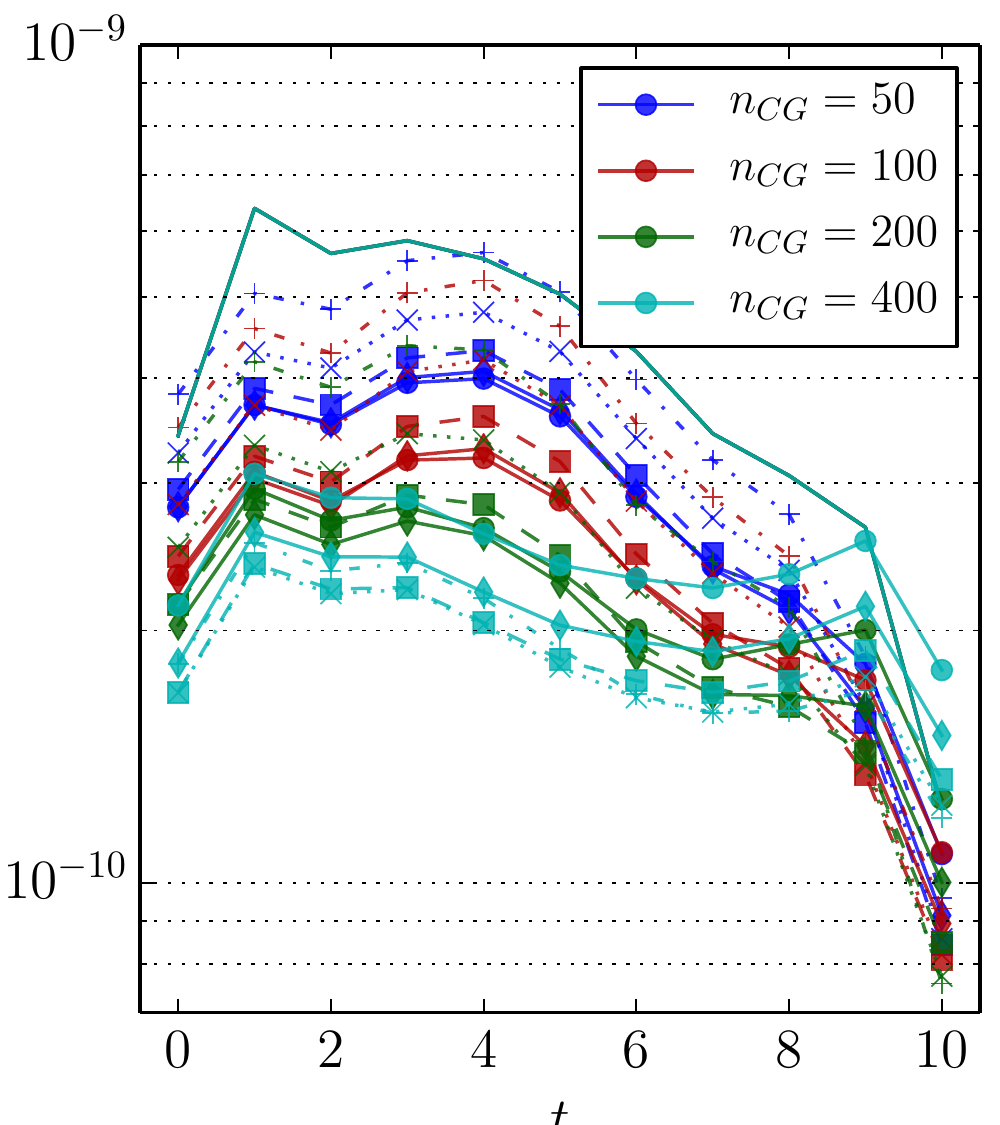}\\
      $\text{Im}\la000|\bar{q}\gamma_3\gamma_5 q|000\ra$
    \end{minipage}~
    \begin{minipage}[T]{.47\textwidth}
      \centering
      \includegraphics[width=\textwidth]{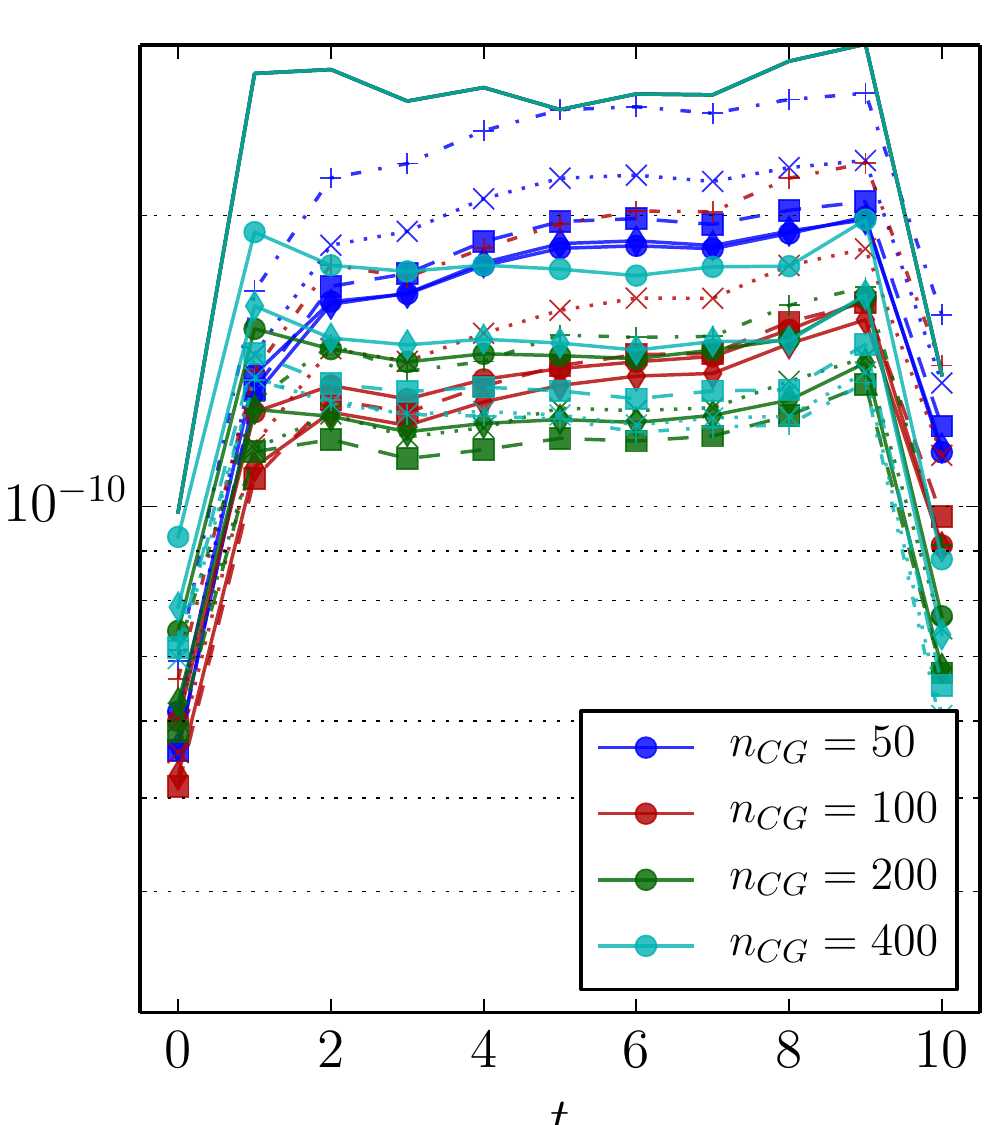}\\
      $\text{Re}\la001|\bar{q}\gamma_4 q|000\ra$
    \end{minipage}\\
    \caption{
      \label{fig:ama-opt-fom}
      Figure of merit for AMA optimization (see Eq.~2.3)
      of the nucleon axial charge (left) and charge radius (right) three-point function plateaus.
      The circles, diamonds, squares, ``x'' and ``+'' correspond to
      $N_\text{approx}/N_\text{exact}=4,8,16,32,64$, respectively, and the solid line corresponds
      to exact-only samples (no AMA).}
  \end{minipage}
\end{figure}

We compute $N_{EV}=500$ lowest eigenmodes for deflation using ARPACK with $n=200$
Chebyshev polynomial acceleration. 
With the range of deflated eigenvalues $\lambda_\text{max}/\lambda_\text{min}\approx10^2$
(see Fig.~\ref{fig:ev-comparison}), the convergence rate of CG is accelerated by a factor of
$\approx10$, which was confirmed by direct tests.
The number of eigenvalues we can use is limited by the total amount of memory
available to a single job.
In fact, using $(3\dots4)\times$ more eigenvectors, which would span the spectrum
between the light and strange quark masses, is expected to improve the AMA 
efficiency significantly~\cite{Blum-Izubuchi-privatecomm}.

In order to apply the AMA effectively, one has to choose the optimal approximation (number of
truncated CG steps $n_\text{CG}$) and ratio of exact and approximate samples 
$N_\text{approx} / N_\text{exact}$. 
Such optimization should take into account the change in the precision of the approximate
samples, as well as their total cost.
We use the product of their stochastic variance and the cost required for their
computation, normalized by the cost of a single exact sample, 
as the figure of merit:
\begin{equation}
\label{eqn:ama-opt-fom}
\text{Cost}_\text{imp}\cdot\text{Var}\big\{\mcO_\text{imp}\big\}
  \sim \Big(1 + \frac{n_\text{CG}^\text{approx}}{n_\text{CG}^\text{exact}}
            \cdot\frac{N_\text{approx}}{N_\text{exact}}\Big)
        \cdot\Big(\text{Var}\big\{\Delta\mcO\big\} 
              + \frac{N_\text{exact}}{N_\text{approx}}\text{Var}\big\{\mcO_\text{approx}\big\}
    \Big).
\end{equation}
Estimates of this FOM for varying combinations of $n_\text{CG}$ and
$N_\text{approx}/N_\text{exact}$ are shown in Fig.~\ref{fig:ama-opt-fom} for the nucleon
three-point functions yielding the axial charge and the charge
radius\footnote{More precisely, the matrix element of the vector current with the minimal non-zero
momentum transfer.}. 
From that, we derive optimal $N_\text{approx}/N_\text{exact}=32$ and
$n_\text{CG}^\text{approx}=400$, which reduce stochastic variance by a factor of $\times(2.5\dots3)$.

Excited states in lattice nucleon matrix elements are a major source of systematic error, and
one has to study the dependence of nucleon-operator three-point functions on the separation
between sources and sinks $T=t_\text{sink} - t_\text{source}$. 
For this initial study, we select $T/a=8,9,10,12$ corresponding to $0.91$, $1.03$, $1.14$, 
and $1.37\text{ fm}$.
Since the statistics are not sufficient for reliable multi-exponential fits, we resort to
simpler methods such as summation (e.g., Ref.~\cite{Mathur:1999uf}) to estimate and reduce
excited-state contributions at this stage.

%%%%%%%%%%%%%%%%%%%%%%%%%%%%%%%%%%%%%%%%%%%%%%%%%%%%%%%%%%%%%%%%%%%%%%%%%%%%%%%
\section{Initial Results\label{sec:results}}

\begin{figure}
  \centering
  \begin{minipage}[T]{.475\textwidth}
    \centering
    \includegraphics[width=\textwidth]{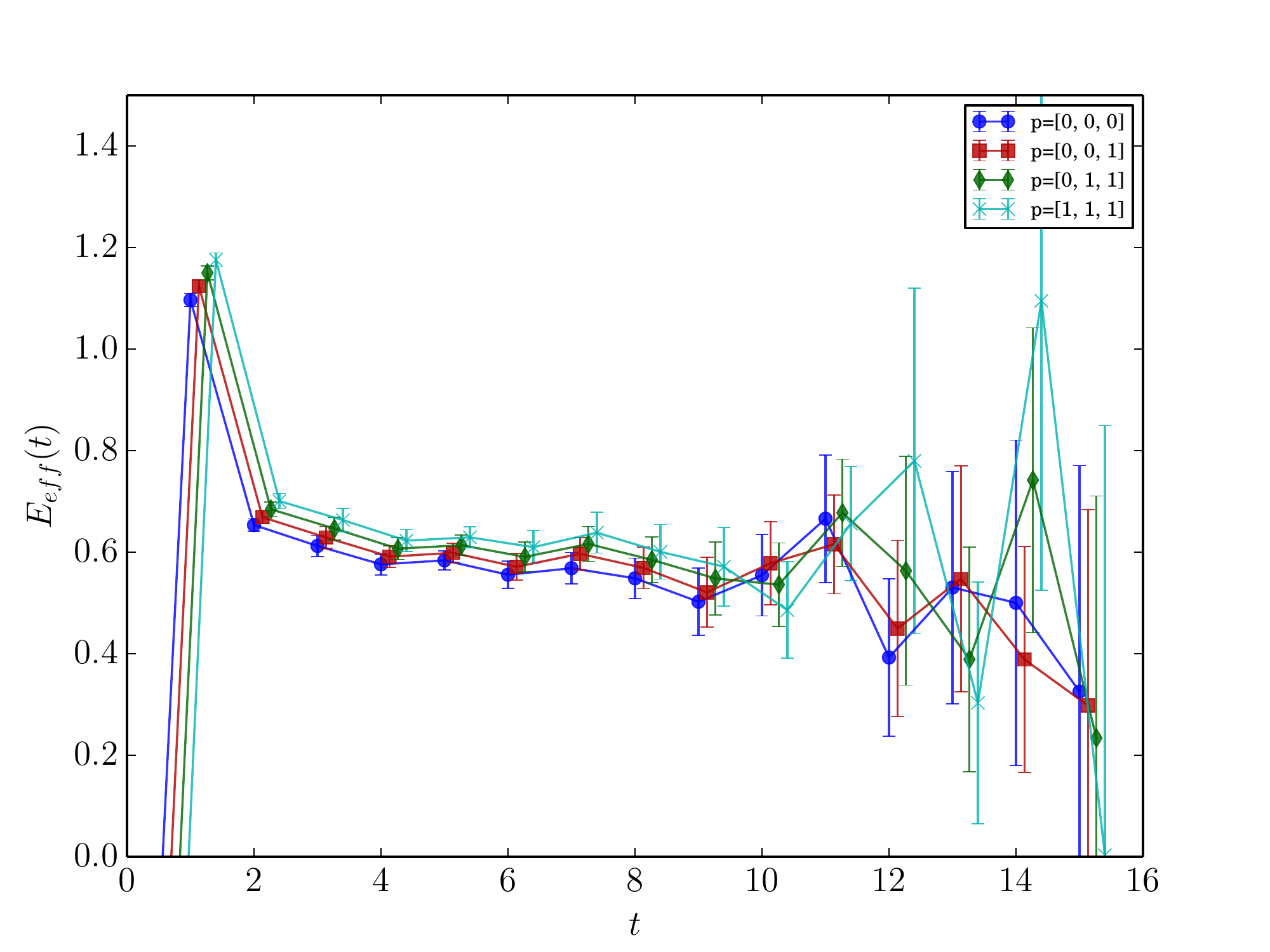}
    \caption{\label{fig:nucleon-meff}Plateaus for nucleon effective energies for the lowest
      momenta. }
  \end{minipage}~
  \hspace{12pt}~
  \begin{minipage}[T]{.475\textwidth}
    \centering
    \includegraphics[width=\textwidth]{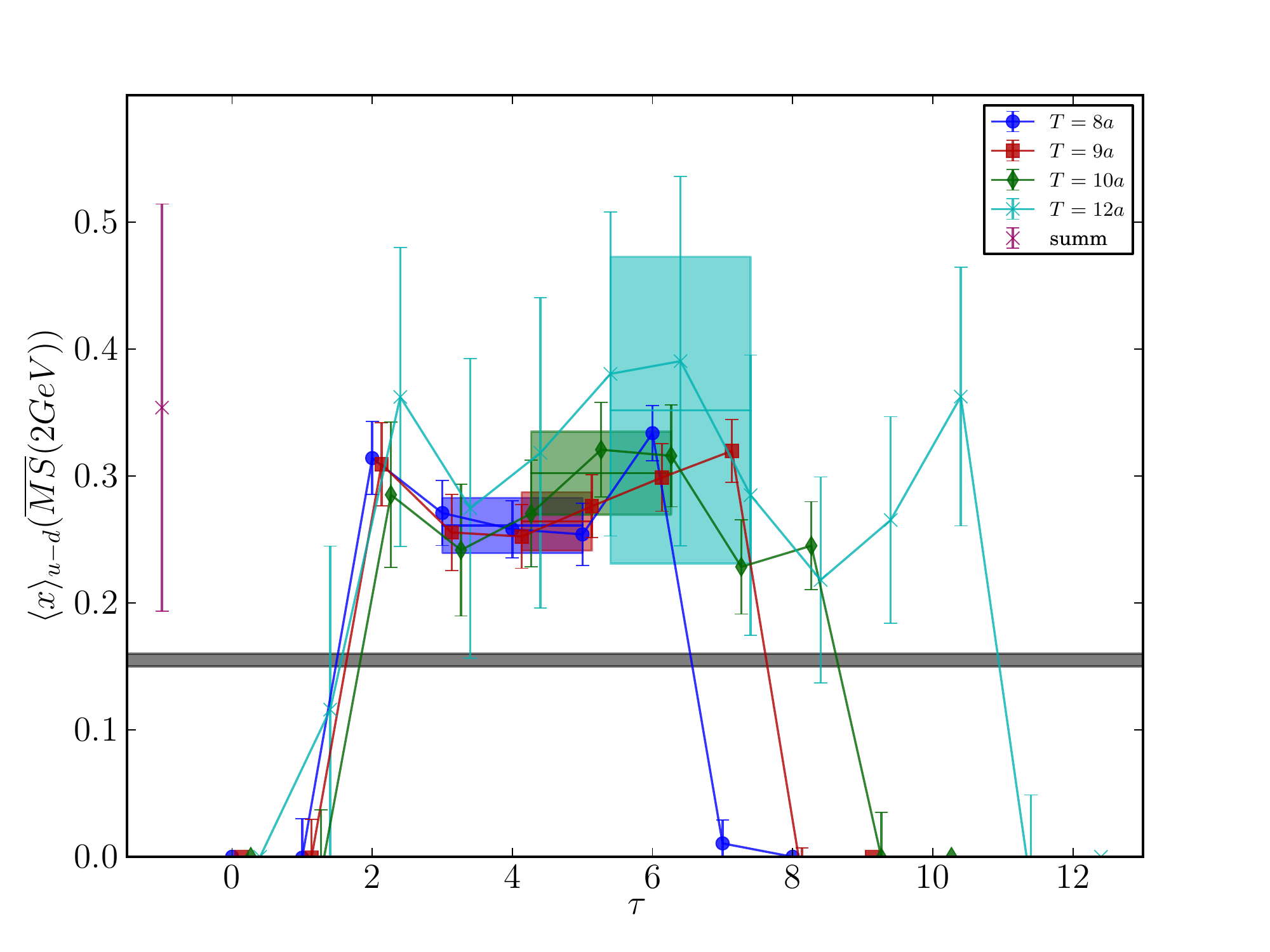}~
    \caption{\label{fig:xv}Plateaus of quark momentum fraction $\la x\ra_{u-d}$. 
%      }
      The renormalization is taken from Ref.~\cite{Aoki:2010xg}.}
  \end{minipage}
\end{figure}

We have analyzed 20 gauge configurations separated by 80 MD steps to minimize autocorrelations;
these configurations are spread evenly over the entire available ensemble except for the first 
640 MD steps necessary for thermalization. 
We compute one exact and 32 approximate samples per configuration.
At this initial stage, we focus on nucleon structure ``benchmark quantities'': 
the nucleon axial charge, Dirac and Pauli isovector radii, isovector magnetic moment 
and quark momentum fraction.
As shown in Fig.~\ref{fig:nucleon-meff}, the effective energies
$E_\text{eff}(t) = \log\big(C_\text{2pt}(t) / C_\text{2pt}(t+1)\big) $ become too noisy after
$t=11$, indicating that the source-sink separation $T/a=12$ is the largest one that can be 
analyzed with the current statistics.
In order to extract nucleon form factors from nucleon matrix elements we use standard methods 
that can be found elsewhere (e.g., see
Refs.~\cite{Yamazaki:2009zq,Aoki:2010xg,Green:2012ud,Green:2014xba}).

\begin{figure}
\centering
\includegraphics[width=.49\textwidth]{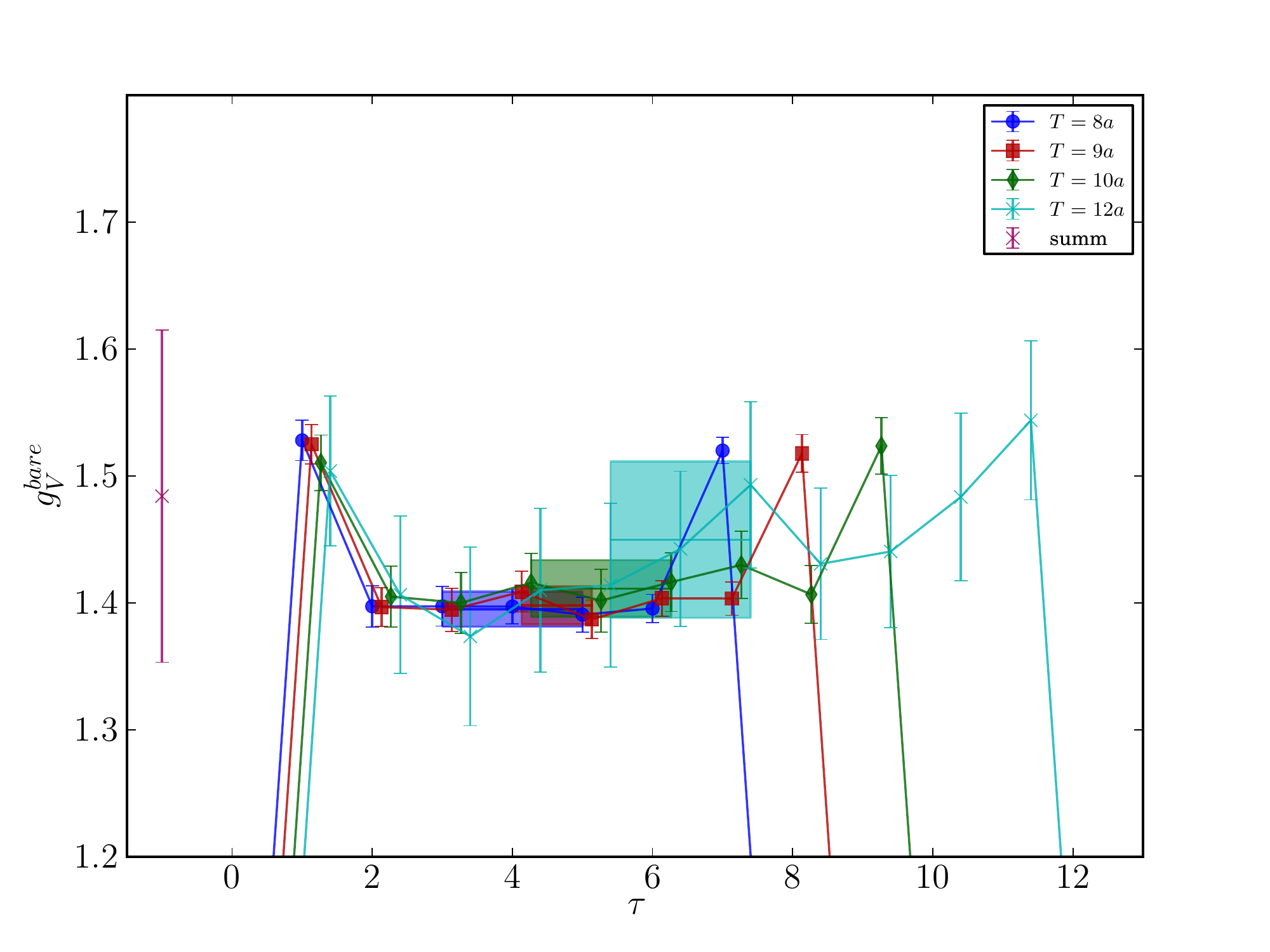}~
\includegraphics[width=.49\textwidth]{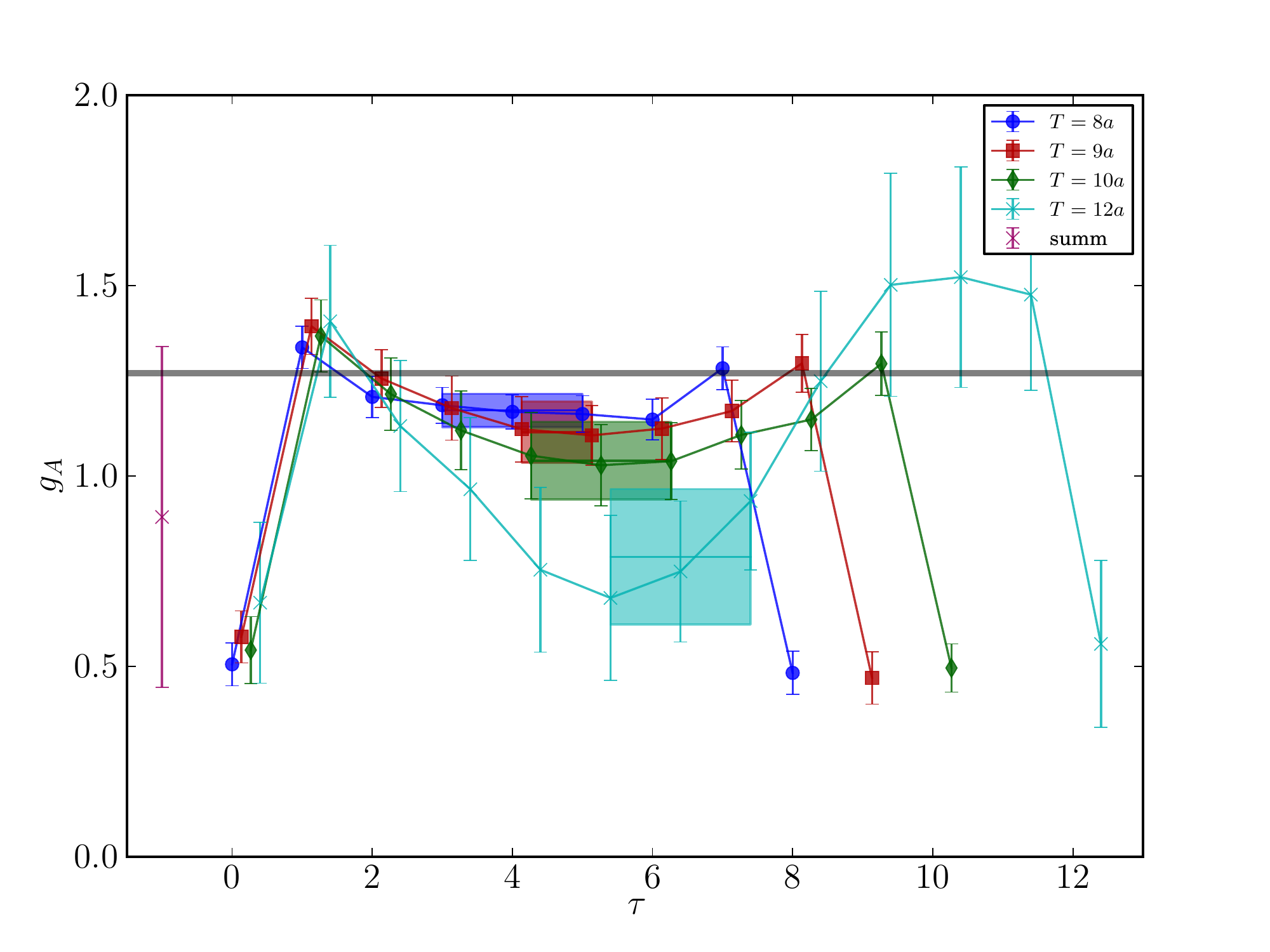}\\
\caption{\label{fig:ga-gv}Plateaus of vector (left) and axial (right) charges of the nucleon.
The leftmost point represents the ``summation'' value.}
\end{figure}

In Figs.~\ref{fig:xv} and~\ref{fig:ga-gv} we show quark momentum fraction, vector charge, 
and axial charge plateaus, their central plateau average values and the values obtained 
with the summation method~\cite{Mathur:1999uf} (applied to data with all the four separations $T$).
Although we have not calculated renormalization factors for the momentum fraction for this ensemble, 
we can use the factors from earlier calculations with Domain Wall fermions with heavier pion masses 
and the identical lattice spacing~\cite{Aoki:2010xg} to convert bare quantities to the 
$\overline{\text{MS}}$ scheme at $2\text{ GeV}$.
The nucleon isovector quark momentum fraction is known to have substantial contributions from excited
states (see, e.g.~\cite{Green:2012ud,Bali:2014gha,Jager:2013kha}), and lattice QCD results
typically overestimate experiment by 30-60\%.
Our initial values for shorter separations also significantly deviate from experiment, while
the longest separation $T/a=12$ has insufficient precision to investigate whether it is 
excited states that cause this systematic effect.
The ``summation'' method yields a value that is consistent with experiment by virtue of its
larger statistical uncertainty.

The vector charge plateaus agree for all separations, and we use the central value at $T=8a$ 
as an approximate renormalization factor for the vector and axial-vector current operators,
$Z_A = Z_V = \big(g_V\big|_{T=8a}\big)^{-1}$.
The value of the axial charge $g_A$ is below the experimental value 
$g_A=1.2723(23)$~\cite{Agashe:2014kda}.
We note that the deviation from the experiment is $(2\ldots3)\sigma$, and it can still be attributed
to a statistical fluctuation.
In addition, the plateaus indicate significant excited state contamination. 
Although the value from the summation method (Fig.~\ref{fig:ga-gv}(right)) 
agrees with experiment, the central plateau values move away from it with increasing separation
$T$, a behavior that was also observed in other studies close to the physical 
point~\cite{Green:2012ud}.
Additional statistics and careful analysis of excited states are required to
understand this phenomenon.

\begin{figure}
\centering
\includegraphics[width=.49\textwidth]{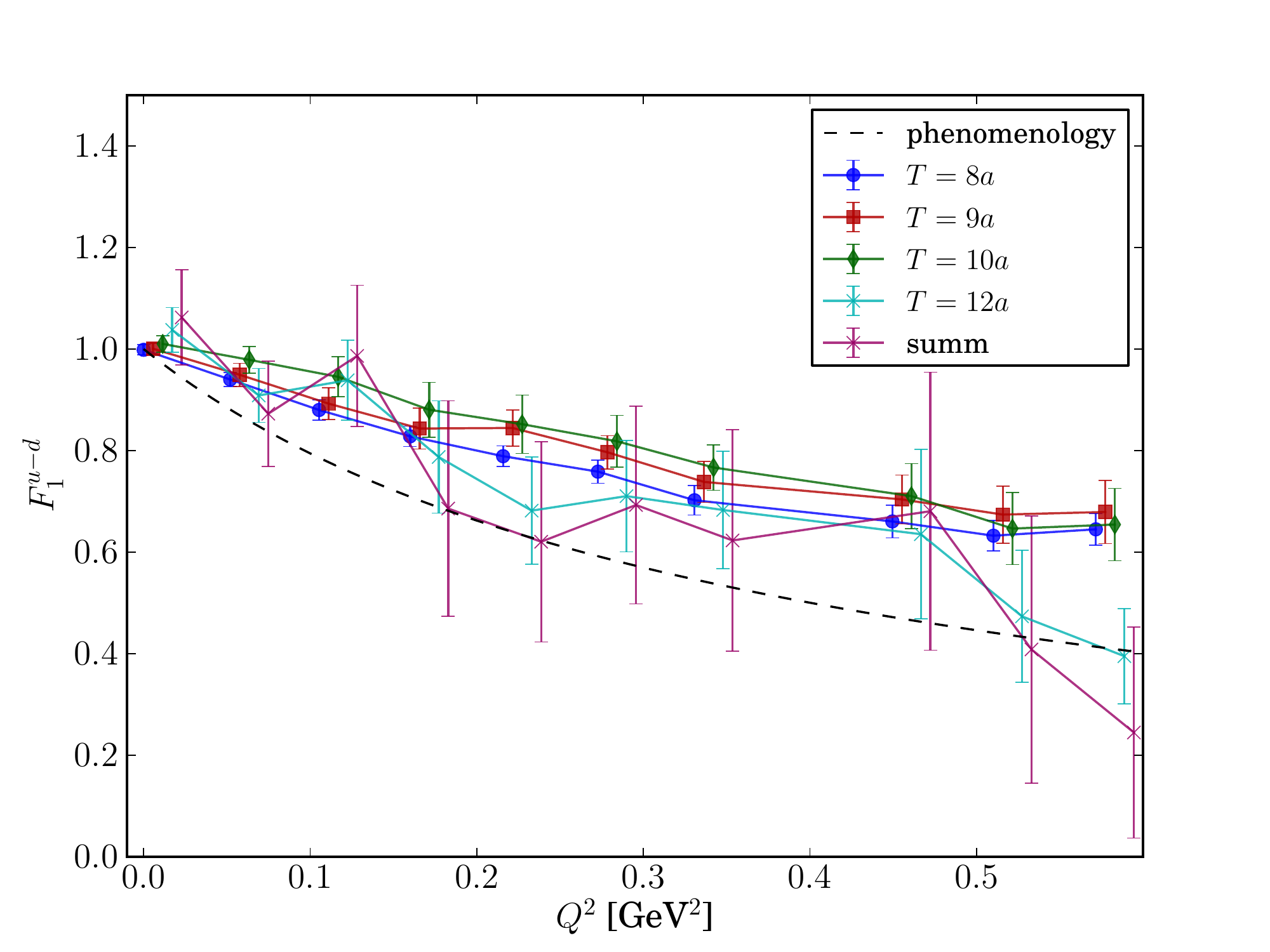}~
\includegraphics[width=.49\textwidth]{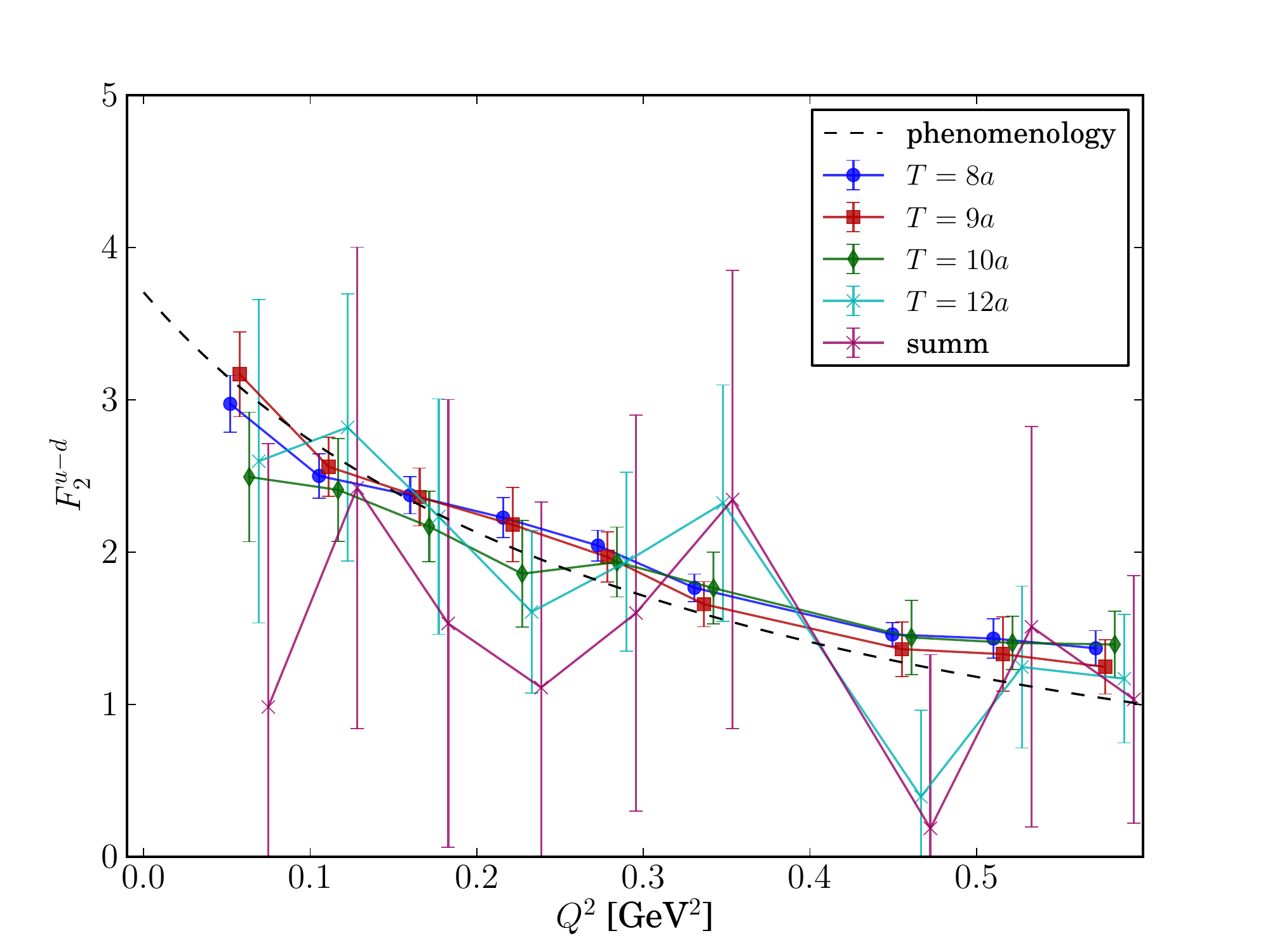}\\
\caption{\label{fig:vec-ff}Vector form factors of the nucleon: isovector Dirac (left) and
  Pauli (right).}
\end{figure}

The statistics are not yet sufficient to analyze the form factors to extract the radii reliably, 
especially with larger separations.
Since no chiral extrapolation is needed, it is more informative to compare the form factors 
directly to experiment.
In Figure~\ref{fig:vec-ff} we show the isovector Dirac and Pauli form factors in the region 
of small momentum transfer $0\le Q^2\lesssim0.6\text{ GeV}^2$,
together with phenomenological fits of experimental data~\cite{Kelly:2004hm}.
The Dirac form factor at small separations $T/a=8,9,10$ deviates from the experiment, 
while the values at the largest separation $T/a=12$, as well as the ``summation'' value,
agree within statistics.
This indicates that the deviations are likely caused by excited state contributions,
in agreement with earlier findings that the isovector Dirac radius is subject to large excited 
state effects~\cite{Green:2012ud}, although more statistics are necessary to make a certain 
conclusion.
The isovector Pauli form factor is in better agreement with phenomenology, although its values 
are even less precise and the ``summation'' method yields nearly 100\% uncertainty.

\begin{figure}
\centering
\includegraphics[width=.49\textwidth]{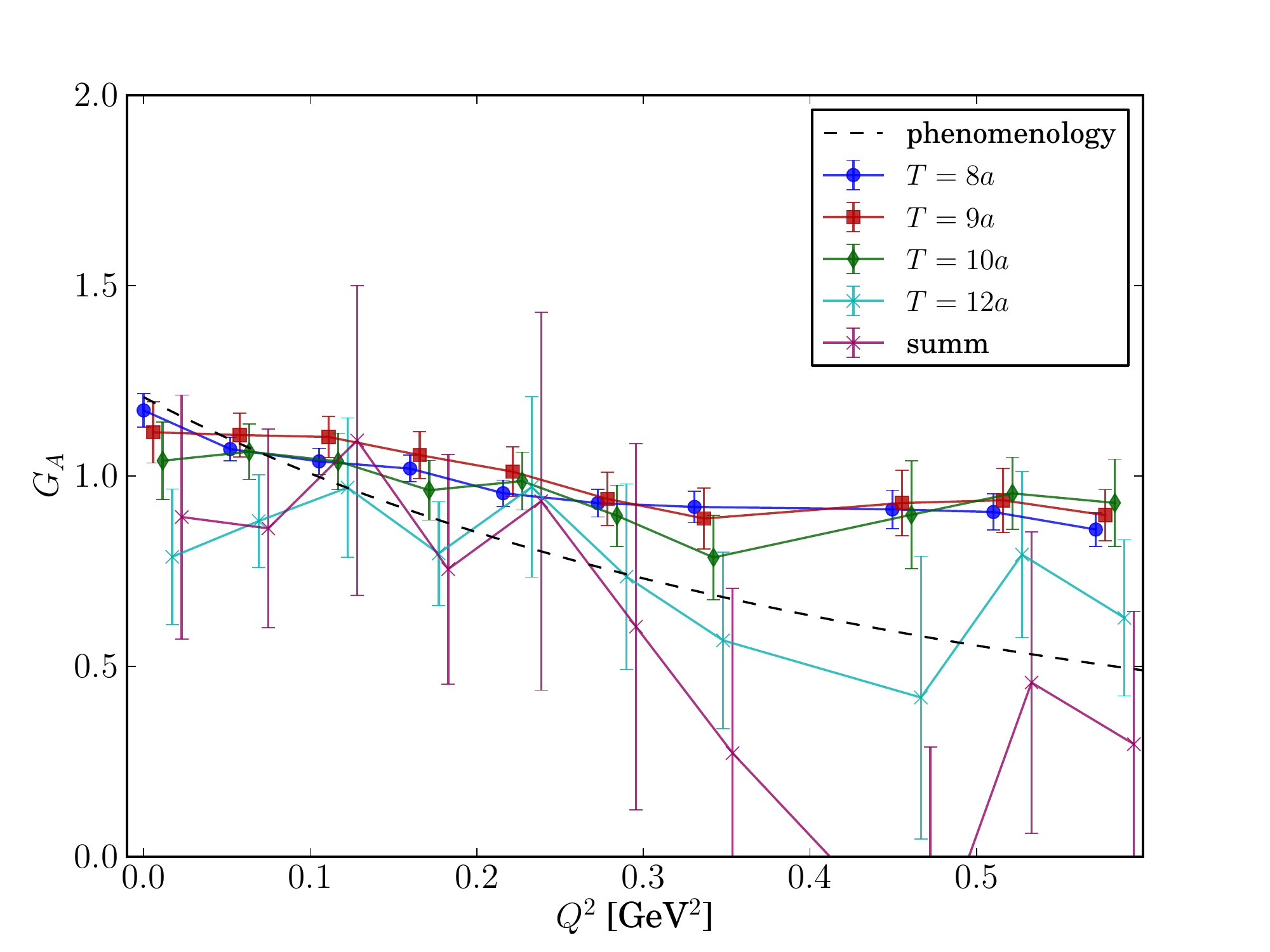}~
\includegraphics[width=.49\textwidth]{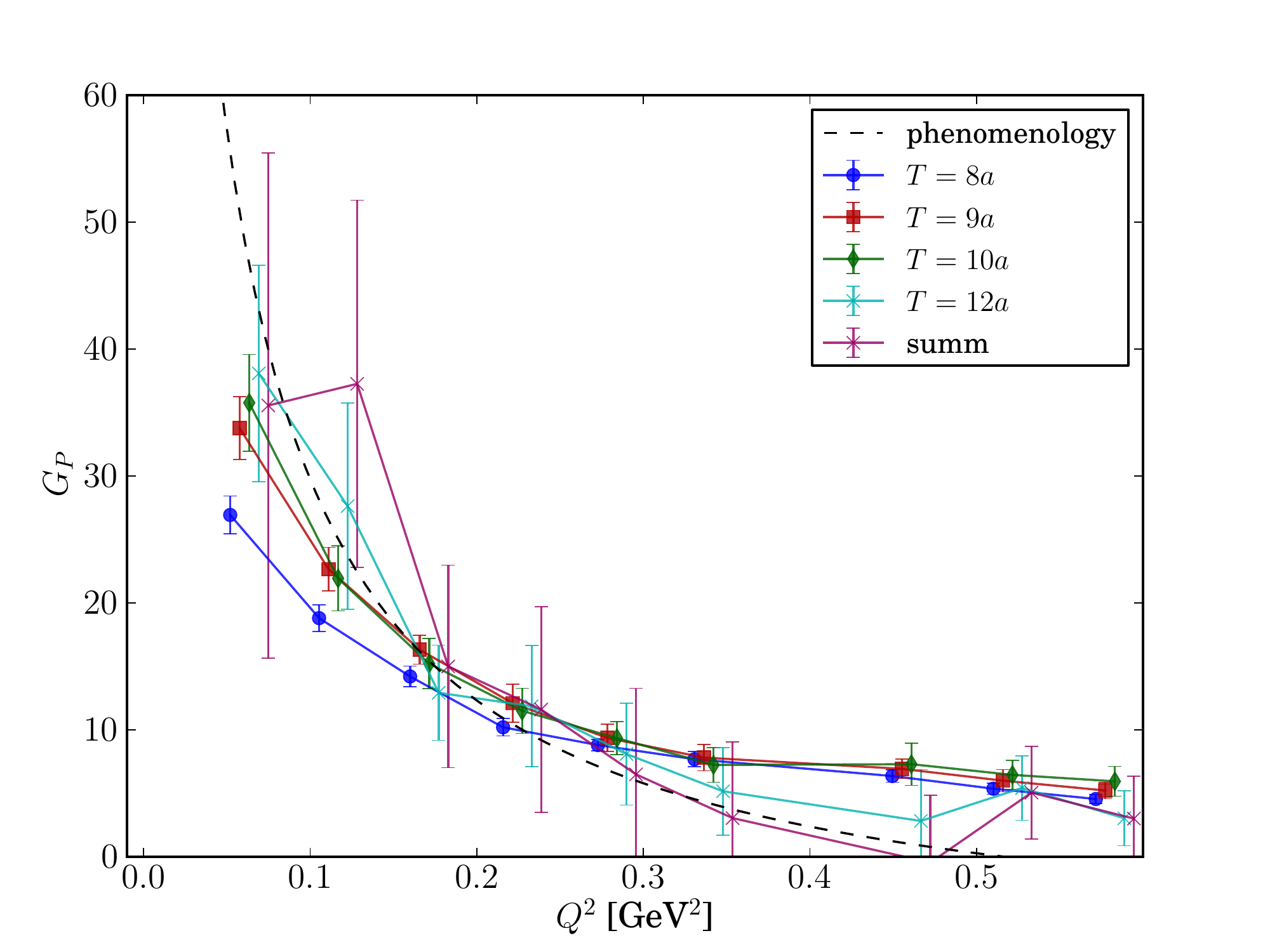}\\
\caption{\label{fig:axial-ff}Axial vector form factors of the nucleon: axial (left) and induced
pseudoscalar (right).}
\end{figure}

The form factors of the axial-vector current are shown in Fig.~\ref{fig:axial-ff}.
The axial form factor $G_A(Q^2)$ is interesting for two reasons.
First, its forward value $G_A(0)=g_A$ is the axial charge that has been discussed above;
second, its $Q^2$-dependence determines the axial radius of the nucleon and plays an important
role in the physics of neutrino scattering and meson production.
The axial radius, defined similarly to the charge radius, is typically underestimated in 
lattice calculations by a factor of 2~\cite{Syritsyn:2011zz}.
From Fig.~\ref{fig:axial-ff}(left) we see that our initial results follow the same pattern: 
the slope of the axial form factor is substantially smaller than experimental fits.
It is not clear yet whether increasing statistics with larger source-sink separations $T/a\ge10$ 
will indicate excited state contamination and suggest a solution to this problem;
however, comparing $T/a=8,9$ and 10 shows very little dependence on the source-sink separation $T/a$.

The induced pseudoscalar form factor $G_P(Q^2)$ (Fig.~\ref{fig:axial-ff}(right)) is 
relevant for low-energy QCD dynamics.
Its value is measured in meson production off nucleons and nuclei, as well as in muon
capture experiments~\cite{Clayton:2009zz}.
Its low-momentum behavior is believed to be governed by the pion pole $\sim(Q^2 + m_\pi^2)^{-1}$
and therefore is very sensitive to the pion mass.
Calculations with the physical pion mass are especially important for this form factor.
Our initial results indicate that there are substantial contributions from excited states,
and as the source-sink separation increases, we observe better agreement with the
phenomenology.

%%%%%%%%%%%%%%%%%%%%%%%%%%%%%%%%%%%%%%%%%%%%%%%%%%%%%%%%%%%%%%%%%%%%%%%%%%%%%%%
\section{Discussion}
Stochastic uncertainty of the initial results that we report indicates that substantially
more statistics will be needed to compute lattice QCD ``benchmark'' quantities with
required precision.
Currently, we plan to quadruple statistics.
Since, for example, our present stochastic uncertainties for the axial charge and the quark momentum
fraction with separation $T=1.14\text{ fm}$ are $\big(\delta g_A\big)_{T/a=10}\approx10\%$
and $\big(\delta \la x\ra_{u-d}\big)_{T/a=10} \approx11\%$, this will reduce their uncertainties 
down to $\approx5\%$ and make reliable analysis of excited states possible.

Although computing resources increase steadily, and the calculation is fully feasible with the
current methodology, we are actively studying approaches to both reduce the cost of computing 
approximate samples and to make them more precise.
An obvious improvement is to simply use more eigenvectors in the CG deflation; 
however, this approach is limited by the size of the 5D eigenvectors ($\approx5.6\text{ TiB}$ 
per configuration).
Other possible directions are using approximations to the M\"obius fermion operator
with shorter $L_5$, multigrid methods, and hierarchical deflation~\cite{Boyle:2014rwa}.

A combination of the improved computing methods and increasing computing resources will allow
us to accomplish a reliable calculation of nucleon structure with chirally symmetric action 
in the near future.

\section{Acknowledgements}
We thank the RBC and UKQCD collaborations for providing us with the gauge configurations.
Computations for this work were carried out on facilities of the USQCD Collaboration, 
which are funded by the Office of Science of the U.S. Department of Energy.
This work was supported by RIKEN Foreign Postdoctoral Researcher Program (S.N.S.) and
the U.S. Department of Energy (DOE) and the Office of Nuclear Physics  under grants
DE-FG02-96ER40965 (M.E.) and  DE-SC0011090 (J.N.).

\iffalse
\bibliographystyle{aip}
\bibliography{syritsyn-lat14-proc}
\else

\input{syritsyn-lat14-proc-bbl}
\fi
%\begin{thebibliography}{99}
%\bibitem{...} 
%....
%
%\end{thebibliography}

\end{document}

%% file: defs.tex
\newcommand{\beq}{\begin{equation}}
\newcommand{\eeq}{\end{equation}}
\newcommand{\bgath}{\begin{gathered}}
\newcommand{\egath}{\end{gathered}}
\newcommand{\bsplit}{\begin{split}}
\newcommand{\esplit}{\end{split}}
\newcommand{\bdm}{\begin{displaymath}}
\newcommand{\edm}{\end{displaymath}}
\newcommand{\beqn}{\begin{eqnarray}}
\newcommand{\eeqn}{\end{eqnarray}}
\newcommand{\bea}[1]{\beq\begin{array}{#1}}
\newcommand{\eea}{\end{array}\eeq}
\newcommand{\bi}{\begin{itemize}}
\newcommand{\ei}{\end{itemize}}
\newcommand{\ben}{\begin{enumerate}}
\newcommand{\een}{\end{enumerate}}
\newcommand{\ba}[1]{\begin{array}{#1}}
\newcommand{\ea}{\end{array}}

\newcommand{\bc}{\begin{center}}
\newcommand{\ec}{\end{center}}
\newcommand{\bfr}{\begin{flushright}}
\newcommand{\efr}{\end{flushright}}
\newcommand{\bfl}{\begin{flushleft}}
\newcommand{\efl}{\end{flushleft}}

\newcommand{\bv}{\begin{verbatim}}
\newcommand{\ev}{\end{verbatim}}

\newcommand{\la}{\langle}
\newcommand{\ra}{\rangle}

\newcommand{\mcO}{{\mathcal O}}